\documentclass[twocolumn,prb,aps,amsmath,amsfonts,amssymb,floatfix,epsf]{revtex4}
\usepackage{graphicx}
\usepackage{epsf}

\begin{document}

\title{Transport in T-shaped ballistic junction}
\author{M. Bek and B. R. Bu{\l}ka}
\affiliation{Institute of Molecular Physics, Polish Academy of Sciences, ul. M. Smoluchowskiego 17, 60-179 Pozna\'{n},
Poland}
\author{J. Wr{\'o}bel}
\affiliation{Institute of Physics, Polish Academy of Sciences, al. Lotnik{\'o}w 32/46, 02-668 Warszawa, Poland}

\begin{abstract}

We present studies of ballistic transport in three terminal T-shaped junction in
a linear and non-linear regime. The floating electrode acts as a scatterer and
modifies the conductance in a direct channel (between source and drain
electrode). In the low voltage limit, the conductance shows the Wigner threshold
effect and the bend resistance. A specific shape of the Wigner singularities can
be changed by applied voltage to the floating electrode as well as by a shift of
the Fermi level. The system also exhibits filtering properties with current
distribution between different modes propagating in the junction. Back action of
current flowing in the direct channel on changes of the voltage in the floating
electrode is considered in the non-linear regime.
\end{abstract}

\maketitle

\section{Introduction} \label{intro}

High-mobility two dimensional electron gas (2DEG) formed at the interface between
GaAs and Al$_x$Ga$_{1-x}$As can be used to make quantum wires (QW's) whose size
is less than the electronic mean free path and electrons can move ballistically
through such structure \cite{Reed}. At low temperatures the phase coherence
length is larger than the size of the device which can lead to a number of
interference effects. One of the examples is the Fano resonance \cite{Fano} which
can take place in QW with donor impurities \cite{Tekman}, coupled to the resonant
cavity \cite{Stone} or to a side attached quantum dot \cite{Kobayashi}.
Interference is important for ballistic transport in narrow constrictions with a
double bend structures \cite{Wu}, with cavities and for many other geometries
\cite{Weisshaar,Laux,Baranger}. Transmission through the such systems shows
series of peaks and dips due to constructive or destructive quantum interference
of electronic waves.

Interference processes are also important in multiterminal devices, e.g. the Fano
resonance caused by localized states in the junction area in the system of three
ballistic wires \cite{Baranger,Schult,Chen}. Another interference effect is bend
resistance, which is caused by destructive interference of outer and inner modes
propagating around the bend area. In the Y-branch switching device one can
observe cusp-like singularities in conductance when new conduction channels open
in a side-attached lead \cite{Worschech}. This effect is called Wigner threshold
effect \cite{Wigner} and is well known in nuclear reactions and scattering of
particles in atomic physics \cite{wignereffect}. In nanostructures the phenomenon
was examined by Baranger \cite{Baranger}, Schult et al \cite{Schult}, Bu{\l}ka
and Tagliacozzo \cite{Bulka}.

Multi-terminal ballistic junctions have attracted much attention for their
potential application in nanoelectronics. Let us mention a few examples such as:
coherently coupled-electron waveguides \cite{Ram}, a frequency-multiplication
device based on a three-terminal ballistic junction \cite{Shorubalko}, diodes as
well as transistors with a very low power consumption \cite{Xu2002}. Three
terminal junctions made purely from carbon nanotubes are very interesting for
fabrication of logic circuits \cite{Papadopoulos,Bandaru,Xu3}. Ferromagnet-normal
conductor multi-terminal system was recently examined by Samuelson and Brataas
\cite{Brat} as a device for quantum state tomography of nonlocal spin
correlations.

Since these interference effects are important for construction of electronic
ballistic devices, we would like to separate and analyze them in details. Firstly
we show that the additional electrode changes scattering conditions which is
manifested in conductance as the Wigner singularities. The shape of these
singularities can be predicted precisely. Secondly, we study filtering properties
of the T-shaped structure in the partial transmission coefficients due to the
presence of the floating electrode. Thirdly, we analyze bend resistance and
mode-matching problem for waves propagating in the T-junction area. Moreover, we
want to study back action of the floating electrode. Although no net current
flows through this electrode, transport in the direct channel (between the source
and drain electrode) is modified. This effect is well pronounced in the
non-linear regime where current flowing in the direct channel induces voltage
changes in the floating electrode \cite{Ram,Xu1,Scho,Xu2}. We also consider how
the Wigner threshold effect and the bend resistance affect electron transport in
the non-linear regime.

The paper is structured as follows. In Sec.II we present a model of three coupled
quantum wires and general derivations of transport properties which are based on
the Landauer-B\"{u}ttiker approach for the ballistic transport \cite{Butt1}
adapted for stripes of atoms. In Sec.III, a multi-channel case is considered in
the linear regime, where we are trying to separate various interference
processes. In Sec.IV we deal with examination of transport properties of
multi-channel T-shaped junction in a non-linear response. The last section is a
summary.

\section{Model of T-shaped junction and current calculations} \label{sec:2}

\begin{figure}[ht]
\includegraphics[width=0.5\textwidth]{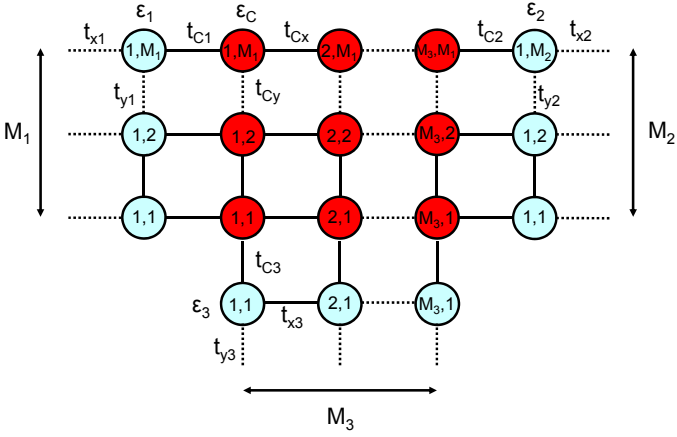}
 \caption{(Color online) Three terminal device which consists of three perfect infinite leads of width $M_i$ ($i=1, 2, 3$) (light blue) and a coupling region (red).}\label{fig1}
\end{figure}

Our three-terminal device is modeled as a system of semi-infinite uniform stripes of atoms of width $M_i a$ ($i=1, 2, 3$ for three quantum wires and $a$ is the lattice constant taken as unity in calculations) connected to a coupling region (see Fig.1). The corresponding Hamiltonian is written in the tight binding form as:
\begin{eqnarray}\label{eq1}
\hat{H}=\sum_{i=1,2,3}\hat{H}_{L_i}+\hat{H}_C+\hat{H}_{C-L},
\end{eqnarray}
where the first term describes electrons in the leads and
\begin{eqnarray}\label{eq2}
\hat{H}_{L_i}=\sum_{x_i,x'_i=1}^{\infty}\sum_{y_i,y'_i=1}^{M_i}[\epsilon_i
a_{x_iy_i}^{\dag}a_{x_iy_i}\nonumber\\
+t_{i}(a_{x_iy_i}^{\dag}a_{x_i,y'_i}+a_{x_iy_i}^{\dag}a_{x'_i,y_i})]
\end{eqnarray}
for $i=1, 2, 3$. Here, the summation is restricted to nearest neighbor sites, $\epsilon_i$ and $t_{i}$ represent the site energies and the hopping integrals, respectively. The spin index are omitted. The term
\begin{eqnarray}\label{eq3}
\hat{H}_C=\sum_{x_c,x'_c=1}^{M_3}\sum_{y_c,y'_c=1}^{M_1}[\epsilon_C a_{x_cy_c}^{\dag}a_{x_cy_c}\nonumber\\
+t_{C}(a_{x_cy_c}^{\dag}a_{x'_c,y_c}+a_{x_cy_c}^{\dag}a_{x_c,y'_c})]
\end{eqnarray}
describes the coupling region, where $\epsilon_C$ and $t_{C}$ denote the site
energies and the hopping integrals. We assume that the coupling region is uniform
(without defects) in order to eliminate scattering which could result in
additional peaks or dips in conductance and complicate our analysis. The last
term in (\ref{eq1})
\begin{eqnarray}\label{eq4}
\hat{H}_{C-L}=t_{C1}\sum_{y_c=y_1=1}^{M_1}(a_{1y_c}^{\dag}a_{1y_1}+\mathrm{h.c.})\nonumber\\
+t_{C2}\sum_{y_c=y_2=1}^{M_2}(a_{M_3y_c}^{\dag}a_{1y_2}+\mathrm{h.c.})\nonumber\\
+t_{C3}\sum_{x_c=y_3=1}^{M_3}(a_{x_c1}^{\dag}a_{1y_3}+\mathrm{h.c.})
\end{eqnarray}
represents coupling between the leads and the central region. Here, $t_{Ci}$ denotes the hopping integral between the coupling region and the first row of atoms in the $i$-th electrode.

To calculate electron transport properties in such device we need to know basic relations between the current/conductance/voltage and the transmission coefficients. In the linear voltage regime one can follow B\"{u}ttiker \cite{Butt1} and write the currents in a multiprobe structure (for temperature
$T=0$) in the form
\begin{eqnarray}\label{eq5}
I_i=\frac{2e^{2}}{h}\sum_{j}T_{ij}(V_{i}-V_{j}),
\end{eqnarray}
where $T_{ij}$ are the transmission coefficients between $i$-th and $j$-th electrode, $V_{j}$ is the voltage in the $j$-th leads. The measured conductance between the $i$-th electrode (source) and $j$-th electrode (drain) when $k$-th electrode is kept floating (no net current is driven through), is
\begin{eqnarray}\label{eq6}
\mathcal{G}_{ij,k}=\frac{2e^{2}}{h}(T_{ij}+\frac{T_{ik}T_{jk}}{T_{ik}+T_{jk}}).
\end{eqnarray}
The second term in the brackets originates from indirect transmission of electrons from $i$-th electrode through $k$-th probe to $j$-th electrode. Using current conservation principle we calculate voltage in the floating electrode
\begin{eqnarray}\label{eq7}
V_k=\frac{T_{ik}V_i+T_{jk}V_j}{T_{ik}+T_{jk}}.
\end{eqnarray}

Let us focus now on the transmission coefficient $T_{ij}$. The scattering theory shows that the coefficient $T_{ij}$ is related to a Green function $G_{ij}$ through transverse wave functions $\chi_{k_i}$ and $\chi_{k_j}$ in the semi-infinite stripes of atoms \cite{Fisher,Ando}
\begin{eqnarray}\label{eq8}
T_{ij}=\sum_{k_i,k_j}T_{k_ik_j},
\end{eqnarray}
\begin{eqnarray}\label{eq9}
T_{k_ik_j}=\upsilon_{k_i}\upsilon_{k_j}|\sum_{y_i=1}^{M_i}\sum_{y_j=1}^{M_j}\chi_{k_i}(y_i)\nonumber\\
\times G_{ij}(x_i,y_j)\chi_{k_j}(y_j)|^{2},
\end{eqnarray}
where $\upsilon_{k_i}=\sqrt{4t_{i}^{2}-(E_i-2t_i\cos k_i)^2}$ and $k_i=\pi n_i/(M_i+1)$ with $n_i=1,...,M_i$. The transverse wave function $\chi_{k_i}$ in the $i$-th electrode is obtained by solving the Schr\"{o}dinger equation for the Hamiltonian (\ref{eq2}) and it is given by
$\chi_{k_i}(y_i)=\sqrt{2/(M_i+1)}\sin(k_iy_i)$. The Green function for the {\it i}-th semi-infinite wire is given by:
\begin{eqnarray}\label{eq10}
G_i(x_i,y_i)=\frac{2}{M_i+1}\sum_{n_y=1}^{M_i}\sin(\frac{\pi n_yx_i}{M_i+1})\nonumber\\
\times g_i(E_i-2t_i\cos(k_i))\sin(\frac{\pi n_yy_i}{M_i+1})\,.
\end{eqnarray}
Here $g_i$ is the one dimensional Green's function for a semi-infinite chain of atoms and is expressed as:
\begin{eqnarray}\label{eq11}
g_i(E) =\left\{ \begin{array}{ll}
\frac{2}{E_i-\sqrt{E_i^{2}-4t_i^{2}}} & \textrm{for $E_i<-2|t_i|$}\\
\frac{2}{E_i+i\sqrt{4t_i^{2}-E_i^{2}}} & \textrm{for $|E_i|<2|t_i|$}\\
\frac{2}{E_i+\sqrt{E_i^{2}-4t_i^{2}}} & \textrm{for $E_i>2|t_i|$,}
\end{array} \right.
\end{eqnarray}
where $E_i=E-\epsilon_i$.

Now, assuming hard walls boundaries in the transverse direction one can derive the Green functions which can be done recursively building up of the system slice by slice from the left side to the right \cite{Ando}. In our analysis we assume that temperature $T=0$ which corresponds to experiments performed at $kT \ll \Delta E$, where $\Delta E$ is the distance between 1D energy levels. For GaAs/GaAlAs quantum wires $\Delta E$ is within the range from 1 meV for split-gate devices up to 10 meV for mesa etched structures \cite{Knop}. Therefore, the limit $kT \ll \Delta E$ is easily obtained in transport measurements.

\section{Linear response regime} \label{sec:3}

\begin{figure}[ht]
\includegraphics[width=0.5\textwidth]{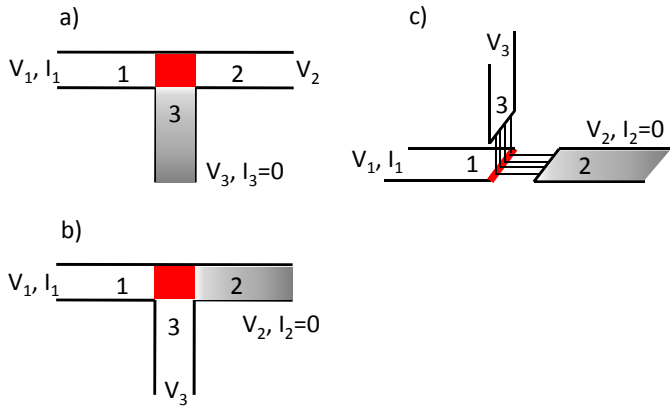}
\caption{(Color online) Schematic presentation of the planar (a, b) and perpendicular (c) configurations of the three terminal system. The coupling region is marked as a (red) square and (red) strip, whereas the floating electrode is marked as shaded area. For example, in the planar straight configuration (a) the current is flowing from 1st to 2nd electrode, whereas 3rd electrodes is kept floating (no a net current is driven through it, $I_3=0$).}\label{fig2}
\end{figure}

\begin{figure}[ht]
\includegraphics[width=0.5\textwidth]{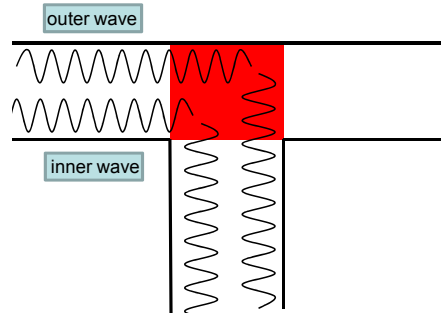}
\caption{(Color online) Schematic view of the bend structure and mode matching problem. The outer and inner modes, with different phase shifts, interfere in a coupling region causing a reduction of transmission.}\label{fig3}
\end{figure}

In this part of our work we would like to consider the situation for a small bias voltage. Two different device configurations are taken into account: a straight and bend one (see Fig.\ref{fig2}) for a planar (a and b) and a perpendicular model (c). Such choice is made to studies back action of the floating electrode and see a role of interference phenomena. First, the planar device is considered for which we expect that interference effects play a relevant role. Electron waves travelling through the bend coupling region (between 1st and 3rd electrode) can be separated as the inner and outer part (Fig.\ref{fig3}). The outer wave travelling through coupling region gains an additional phase with respect to the inner one which has a shorter optical path. Matching of wave modes in the coupling region and their destructive interference lead to reduction of transmission and to bend resistance. One can expect that the bend resistance increases with an increase of the phase difference between the outer and inner modes, with an increase of the wave vector of the transmitted wave.

The second interference phenomenon is the Wigner threshold effect which can be pronounced in energy range when a new channel is open in the lead \cite{Wigner}. The threshold effect originates from conservation of the flux. It has the quantum mechanical origin and exhibits specific threshold singularities in the scattering matrix on both sides of a threshold energy. Those two effects can appear in the same energy range and the conductance plot can be complex. To separate the effects we use the second model with the perpendicular configuration of electrodes where the coupling region is reduced to the single strip (see Fig.\ref{fig2}c). In this model there is no phase difference between outer and inner wave modes in the coupling region, and therefore, neither mode matching problem nor bend resistance take place.

\subsection{Straight configuration}

\begin{figure}[ht]
\includegraphics[width=0.5\textwidth]{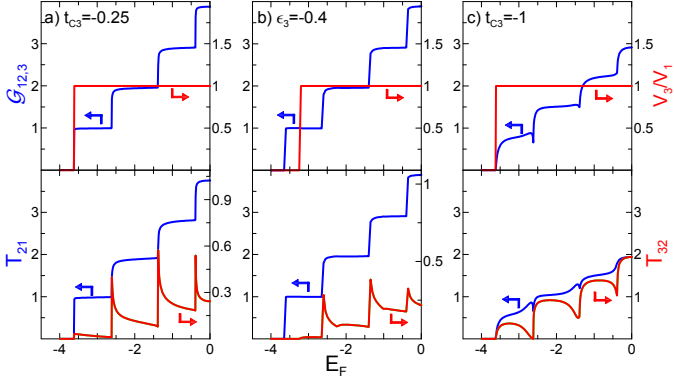}
\caption{(Color online) Results for the planar device in the straight configuration (with the 3rd electrode left floating): conductance $\mathcal{G}_{12,3}$ (blue), voltage $V_3$ (red) in the floating electrode and (bottom panel) transmission coefficients $T_{21}$ (blue) and $T_{32}=T_{31}$ (red) as a function of Fermi energy $E_F$ for (a) $t_{C3}=-0.25$, $\epsilon_{3}=0$, (b) $t_{C3}=-0.25$, $\epsilon_{3}=-0.4$ and (c) $t_{C3}=-1$, $\epsilon_{3}=0$. For all cases: $M_1=M_2=M_3=4$, $\epsilon_C=0$. Threshold energy $E_F^{th1}=-2.62$, $E_F^{th2}=-1.48$, $E_F^{th3}=-0.48$.}\label{fig4}
\end{figure}

Let us first consider the planar device in the straight configuration presented in Fig.\ref{fig2}a. The current is driven through the direct channel (from the 1st to 2nd electrode) when the 3rd electrode (the side channel) is floating and weakly connected to the coupling region. In this situation we can suppose that transmission through the direct channel as a function of the Fermi energy $E_F$ is similar to that which one observes through quantum wire with conductance plateaux \cite{Wees}. The 3rd electrode acts as an inelastic scatterer which modifies the conductance curve. The situation for a weak coupling with 3rd electrode is presented in Fig.\ref{fig4}a. The transmission $T_{21}$ steps are well pronounced and quantized at $2e^2/h$ and its multiple. The transmission $T_{32}$ shows a sharp cusp when a new channel opens at the threshold energy $E_F^{th}$. These changes of transmission are due to the Wigner threshold effect. However, this effect is difficult to see in the conductance $\mathcal{G}_{12,3}$ [see Eq.(\ref{eq6}) and the blue curve in the upper panel of Fig.\ref{fig4}a], because the transmissions $T_{21}$ is much greater than $T_{32}=T_{31}$ and masks its influence.

The results in Fig.\ref{fig4}b are for the case with the site energy in the floating electrode shifted by applying an additional potential $\epsilon_{3}=eV_{G3}=-0.4$ causing the side channel to be opened as the second one after the direct channel. The conductance $\mathcal{G}_{12,3}$ is much like the same as in the previous case, but the transmission $T_{32}=T_{31}$ to the floating electrode manifests new features. When the new side channels opens at $E_F=-3.2$, $E_F=-2.3$ and $E_F=-1$, we observe the small drop of the transmission $T_{21}$ due to the bend resistance (the effect is not pronounced here). When the new channel opens between the source and the drain, we see that $T_{31}$ rises drastically forming a cusp.

The Wigner cusps disappear for a strong coupling of 3rd electrode when scattering processes are strong (see Fig.4c). In this case, the 3rd electrode acts as an inelastic scatterer, which breaks coherent electron transport in the main channel. The scattering processes significantly lower the conductance plateau below $2e^2/h$ and its multiple. The steps in the transmission $T_{21}$ are significantly blurred and show small dips close to the threshold energies. $T_{32}$ shows wide dips, in contrast to the sharp peaks which appeared for the weak coupling case (Fig.\ref{fig4}a). These dips are manifestation of bend resistance which will be analyzed in details later.

There are four possible shapes of the Wigner threshold singularities such as: upward cusp, downward cusp and two saddle-like singularities \cite{Wigner}. Opening of a new reaction channel in the system of colliding particles leads to appearing of threshold singularities in the differential scattering cross-section. The scattering amplitude near the reaction threshold can be written as $f_t(\theta,E)=f_t(\theta)-(k_t/4\pi)A\sqrt{E-E_t}e^{2i\delta}$, where $f_t(\theta)$ is scattering amplitude at threshold energy $E_t$, $\theta$ is a
scattering angle, \linebreak $k_t = \sqrt{2mE_t}/\hbar$ is the wave number of a colliding particle and $\delta$ is the phase shift of the scattered particle. The differential cross-section is described by:
\begin{eqnarray}\label{eq12}
\frac{d\sigma}{do}=|f_t(\theta)|^2-\frac{k_t}{2\pi}A\sqrt{E_t-E}\;
\Re[f_t(\theta)e^{-2i\delta}]\nonumber\\ \textrm{for $E<E_t$},\\
\label{eq13}
\frac{d\sigma}{do}=|f_t(\theta)|^2+\frac{k_t}{2\pi} A\sqrt{E-E_t}\;\Im[f_t(\theta)e^{-2i\delta}]\nonumber\\
 \textrm{for $E>E_t$}.
\end{eqnarray}
Close to the threshold energy $d\sigma/do$ is a linear function of $\sqrt{|E-E_t|}$ with different slopes above and below the threshold. If the scattering amplitude has the form $f_t(\theta)=|f_t|e^{i\alpha(\theta)}$ (where $\alpha(\theta)$ is an unknown phase of scattered particle), then the shape of the differential cross-section is dependent on whether the angle $2\delta-\alpha$ is in the first, second, third or fourth quadrant.

The Wigner cusps should be also pronounced in coherent transport in nanostructures. The origin of the singularity is interplay of the flux conservation and the interference processes between incoming and reflected waves when a channel in a nearby side terminal is opened. Calculations \cite{Schult} in crossed quantum wires or in four terminal junctions showed singularities in the scattering cross section. The simplest system is T-structure of ballistic wires described by a model of three semi-infinite chains of atoms. Using the derivations in the Section \ref{sec:2} one can determine the conductance $\mathcal{G}_{12,3}$ in the detector (when the source-drain voltage is applied to quantum wire 1 and 2, whereas wire 3 is left unbiased) as:
\begin{eqnarray}\label{eq14}
\mathcal{G}_{12,3}=\frac{2e^2}{h}\upsilon_1\upsilon_2 t_{01}^2 t_{02}^2|g_1|^2|g_2|^2|G_{00}|^2\nonumber\\
\times[1+\frac{\upsilon_3
t_{03}^2|g_3|^2}{\upsilon_1t_{01}^2|g_1|^2+\upsilon_2
t_{02}^2|g_2|^2}],
\end{eqnarray}
where
\begin{eqnarray}\label{eq15}
G_{00}=\frac{1}{E-\epsilon_0-t_{01}^2g_1-t_{02}^2g_2-t_{03}^2g_3}
\end{eqnarray}
describes the Green function of the central site at which all three wires are connected.

When the electron energy $E_i=E_F-\epsilon_i$ enters the energy band in the wire (i.e. $|E_i|\leq 2|t_i|$) then the coefficient $\upsilon_i=\sqrt{4t_i^2-E_i^2}$ and vanishes in other cases. An additional gate voltage $eV_{G3}$ applied to the 3rd wire shifts its site energy by $\epsilon_3\rightarrow \epsilon_3+eV_{G3}$. By moving $E_3$ (by $eV_{G3}$ or $E_F$) below the bottom of the conduction band the coefficient $\upsilon_3$ vanishes at the threshold energy $E_t=-2|t_3|$. The second term in brackets in (\ref{eq14}) describes an indirect transmission via the floating electrode [see also the second term in (\ref{eq6})] and it vanishes below the threshold $E_t$.

To examine the Wigner singularities one has to expand $\mathcal{G}_{12,3}$, Eq.(\ref{eq14}), in a series with respect to
\begin{displaymath}
\nu_3=\sqrt{E_3^2-4t^2_3}
\end{displaymath}
and
\begin{displaymath}
\upsilon_3=\sqrt{4t_3^2-E_3^2}
\end{displaymath}
for the energies $E_3$ below and above the threshold energy $E_t$, respectively:
\begin{eqnarray}\label{eq16}
\mathcal{G}_{12,3}=\mathcal{G}_{12,3}|^{0^-}[1+\frac{4t_{03}^2\;\nu_3}{(E_F-E_t)^2}\Re[G_{00}|^{0^-}]],
\end{eqnarray}
\begin{eqnarray}\label{eq17}
\mathcal{G}_{12,3}=\mathcal{G}_{12,3}|^{0^+}[1+\frac{4t_{03}^2\;\upsilon_3}
{(E_F-E_t)^2}(\Im[G_{00}|^{0^+}]+
\nonumber\\
+\frac{1}{t_{01}^2|g_1|^2\upsilon_1+t_{02}^2|g_2|^2\upsilon_2})],
\end{eqnarray}
where $\mathcal{G}_{12,3}|^{0^{\mp}}$ is the conductance between 1st and 2nd wire, when 3rd wire is floating, below ($-$) and above ($+$) the energy threshold. If we notice that the Green function for a semi-infinite strip of atoms can be written as $g_i=e^{ik_i}/t_i$, we see that equations (\ref{eq16}) and (\ref{eq17}) are similar with (\ref{eq12}) and (\ref{eq13}), respectively. Both pairs of equations describe a singular behaviour of the differential cross-section and the conductance when the phase of colliding particles changes. Below the threshold, the conductance $\mathcal{G}_{12,3}$ (Eq.\ref{eq16}) describes the elastic processes - similar to the cross-section (Eq.\ref{eq12}). Above the threshold Eq.\ref{eq17} presents the elastic as well as the inelastic contributions to the conductance.

In the above case we have assumed that the system consists of three semi-infinite chains of atoms. Our numerical calculations are performed for the T-device of strips of atoms with a different width $M_i>1$. In this case one has many travelling modes with different perpendicular wave numbers. Such device is more realistic, in which one can observe bend resistance as well as different shapes of the Wigner singularities.

\begin{figure}[ht]
\includegraphics[width=0.5\textwidth]{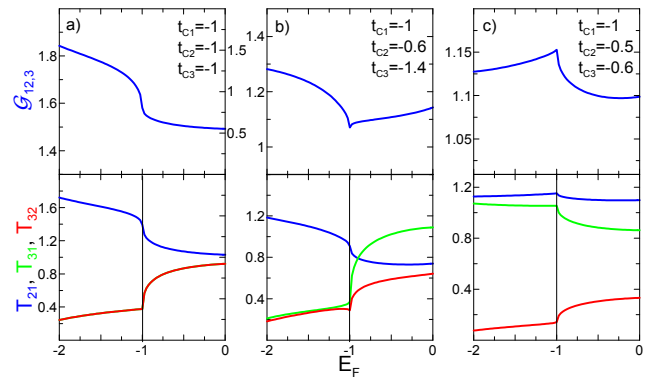}
\caption{(Color online) Three (of four) Wigner threshold singularities which appear in conductance $\mathcal{G}_{12,3}$ as a function of Fermi energy. A thin vertical line indicates threshold energy. For all plots $eV_{G3}=-1$, $\epsilon_{Ci}=0$, $\epsilon_C=0$ and $M_i=3$ ($i=1,2,3$). The plots were made for the device configuration presented in Fig.\ref{fig2}a).}\label{fig5}
\end{figure}

Fig.\ref{fig5} presents different forms of the Wigner cusps for the T-device in the straight configuration with the atomic strips of the width $M_i=3$ and with the potential $V_{G3}$ applied to the 3rd electrode, when the Fermi energy is varied. The conductance $\mathcal{G}_{12,3}$ forms saddle-like singularity (Fig.\ref{fig5}a) for the system with the same coupling with the electrodes ($t_{C1}=t_{C2}=t_{C3}=-1$). The Wigner effect appears at $E_F=-1$ because the conduction band in 3rd electrode is shifted by $eV_{G3}=-1$. Transmission in the direct channel $T_{21}$ shows a saddle-point cusp, whereas $T_{31}=T_{32}$ has an inverse saddle-point dependence (see the bottom plots in Fig.5a). The conductance $\mathcal{G}_{12,3}$ is a composition of the direct and indirect transmission (see Eq.\ref{eq6}) and therefore $T_{21}$ plays a major role. We observe downward and upward cusps in Fig.\ref{fig5}b and Fig.\ref{fig5}c for different couplings to the wires ($t_{C1}\neq t_{C2}\neq t_{C3}$). The shape of the singularities depends on relatively coupling to the electrodes. For a strong coupling to the 3rd electrode, $|t_{C3}|>|t_{C2}|$, (see Fig.5b) the transmissions $T_{21}$,
$T_{31}$, $T_{32}$ have similar shapes as in the previous case (with the saddle-point and inverse saddle-point dependence). However, above the threshold $T_{31}$ becomes dominating which results $\mathcal{G}_{12,3}$ in the downward cusp. For weak coupling to the floating electrode ($|t_{C1}|>|t_{C2}|\approx |t_{C3}|$ (see Fig.\ref{fig5}c) the transmissions $T_{21}$ and $T_{31}$ are dominating and show the upward cusp. That is why, the conductance $\mathcal{G}_{12,3}$ also exhibits the upward cusp.

\begin{figure}[ht]
\includegraphics[width=0.45\textwidth]{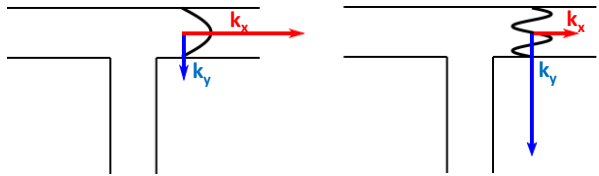}
\caption{Schematic view of the TBJ structure and the length of the forward-directed $k_x$ and the transverse-directed $k_y$ wave vector for the low-lying (left panel) and the high-lying mods (right panel).}\label{fig6}
\end{figure}

In the T-shaped structures one can find another interesting phenomenon - the filtering of electron waves \cite{Baranger}. Having a whole pool of incoming modes we can observe how the junction changes the distribution of the electrons among the modes of the waveguide. The low lying mode corresponds to an electron with a large forward-directed wave vector $k_x$ and the small transverse-directed wave vector $k_y$, while high lying modes correspond to a small forward and long transverse wave vector (see Fig.\ref{fig6}). According to Eq.(\ref{eq8}) the transmission coefficient $T_{ij}$ between the $i$-th and $j$-th lead is a sum of partial transmissions $T_{k_ik_j}$, where $k_i=\pi n_i/(M_i+1)$ corresponds to the wave vector of transverse modes and $n_i$ is the mode number. In order to study filtering we consider the transmission $\sum_{k_j} T_{n_i,k_j}$ to the $n_i$ mode in the $i$-electrode which takes into account the whole pool of incoming wave modes from the $j$-th electrode.

\begin{figure}[ht]
\includegraphics[width=0.5\textwidth]{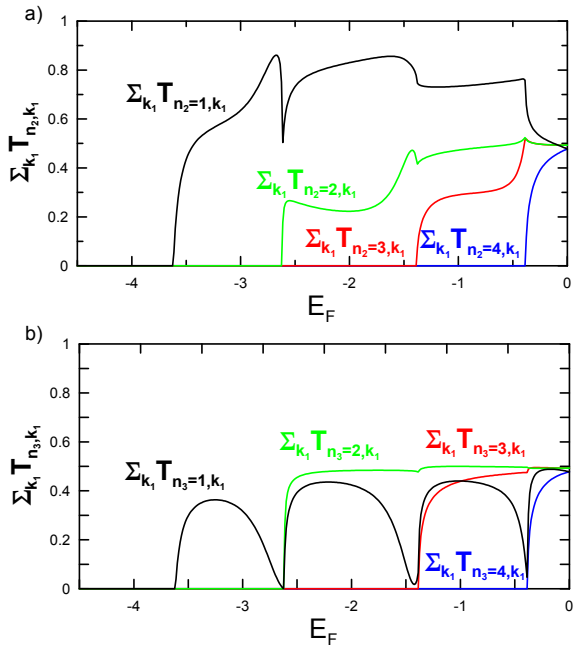}
\caption{ Partial transmissions $\sum_{k_1} T_{n_2,k_1}$ between 1st and 2nd electrode (a) and $\sum_{k_1} T_{n_3,k_1}$ between 1st and 3rd electrode (b) as a function of Fermi energy $E_{F}$. For all plots $t_{Ci}=-1$ ($i=1,2,3$), $\epsilon_C=0$ and $M_1=M_2=M_3=4$.}\label{fig7}
\end{figure}

Fig.\ref{fig7}a and b present filtering properties for different modes propagating  from the second and the third electrode to the first electrode. The lowest lying mode
\begin{displaymath}
\sum_{k_1} T_{n_2=1,k_1}
\end{displaymath}
is transmitted through the direct channel with high probability because its transverse directed wave vector is small. This travelling wave propagates straight through the wire and there is small chance to turn the corner to the floating electrode. With increasing mode multiplicity the chance that the electron waves turn around the bend (between 1st and 3rd electrode) increases and the higher modes have reduced transmission. When new conduction channels are open simultaneously in all three electrodes, inelastic processes modify the partial transmission and its shape is different from step-like behaviour. We observe an increase of the transmission probability of the modes with higher multiplicity
\begin{displaymath}
\sum_{k_1} T_{n_2=1,k_1}>\sum_{k_1} T_{n_2=2,k_1}>....
\end{displaymath}
In Fig.\ref{fig7}b are presented partial transmissions for the bend configuration (between the first and the second and 3rd electrode). The filtering properties are not well pronounced. For example, the transmission probability $\sum_{k_1}T_{n_3=1,k_1}$ for the first mode can be lower than $\sum_{k_1} T_{n_3=2,k_1}$ and $\sum_{k_1} T_{n_3=3,k_1}$ for the second and third mode. This shows an influence of bend resistance in the coupling region which will be described later.

\subsection{Bend configuration}

\begin{figure}[ht]
\includegraphics[width=0.5\textwidth]{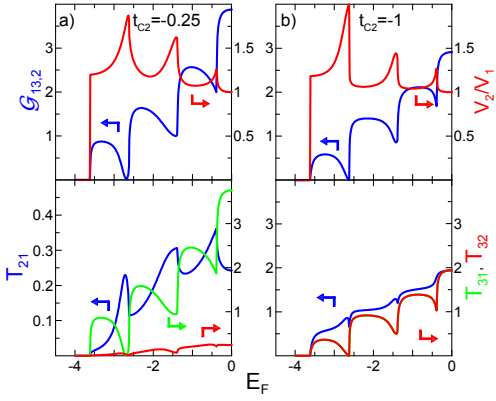}
\caption{(Color online) Results for the planar device in the bend configuration (with the 2nd electrode left floating): conductance $\mathcal{G}_{13,2}$ (blue), voltage $V_2$ (red) in the floating electrode and (bottom panel) transmission coefficients $T_{21}$ (blue), $T_{31}$ (green) and $T_{32}$ (red) as a function of Fermi energy $E_F$ for (a) $t_{C2}=-0.25$ and (b) $t_{C2}=-1$. For all cases: $M_1=M_2=M_3=4$, $\epsilon_C=0$, $\epsilon_{i}=0$. Threshold energy $E_F^{th1}=-2.62$, $E_F^{th2}=-1.48$, $E_F^{th3}=-0.48$.}\label{fig8}
\end{figure}

Now it is worth considering the linear transport in the bend configuration presented in Fig.\ref{fig2}b. The current is driven between 1st and 3rd electrode when the latter one is left floating. The results for weak coupling are presented in Fig.\ref{fig8}a. In this situation the symmetry of the device is broken (due to different couplings to the electrodes, $t_{C1}=t_{C3}\neq t_{C2}$). It manifests itself in differences in the transmission probabilities $T_{31}$ and $T_{32}$. The transmission $T_{31}$ to the floating electrode shows deeps at every threshold energy which are caused by bend resistance. The transmission $T_{21}$ shows sharp peaks in contrast to the indirect transmission $T_{32}$. This results from a different mechanism, namely from the Wigner threshold effect. Bend resistance reduces the flux of particles transmitted from the 1st to the 3rd electrode and  a significant part of electrons is injected to the floating electrode 2.

In accordance with Eq.\ref{eq6} the major role in conductance $\mathcal{G}_{13,2}$ plays transmission $T_{31}$ with large dips on its characteristics. On the contrary to the previous case presented in Fig.\ref{fig4} the voltage $V_2$ in the floating electrode does not remain constant, but now oscillates and shows large peaks at every threshold energy (see the red curves in the top panel in Fig.\ref{fig8}). This kind of voltage graph results from the flux conservation of transmitted particles. It is described more precisely and quantitatively by Eq.\ref{eq7}. The large peaks in $V_2$ result from the differences in the transmission coefficients $T_{21}$ and $T_{32}$ to the floating electrode 2 (see the corresponding plots in the bottom panel in Fig.\ref{fig8}).

For equal couplings of the electrodes $t_{Ci}=-1$ ($i=1, 2, 3$) (see Fig.\ref{fig8}b)  the transmission $T_{21}$ shows small peaks whenever the mod matching problem takes place. The indirect transmissions coefficients are equal $T_{31}=T_{32}$ and shows deeps at every threshold energy due to bend resistance. Far from the scattering region the waves are planar waves, but their inner and outer parts have different optical path lengths in the bending region. There is a mode matching problem for the incoming and outgoing waves. This effect dominates in the bend configuration.

The conductance $\mathcal{G}_{13,2}$ shows small plateaux, but is still suppressed  below the quantum of conductance $2e^2/h$ and its multiplicity. Moreover, it shows well pronounced dips in its characteristics because of bend resistance. The voltage $V_2$ calculated in the floating electrode oscillates, because transmissions to the floating electrode are unequal ($T_{21} \neq T_{32}$) and $V_2$ shows sharp curvature with large peaks whenever the channel is closing. It tells us that electrons are strongly reflected from the 2nd electrode.

\begin{figure}[ht]
\includegraphics[width=0.5\textwidth]{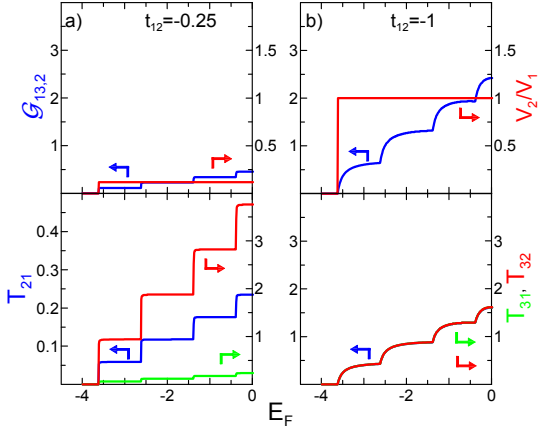}
\caption{(Color online) Results for the perpendicular device in the bend configuration (with the 2nd electrode left floating): conductance $\mathcal{G}_{13,2}$ (blue), voltage $V_2$ (red) in the floating electrode and (bottom panel) transmission coefficients $T_{21}$ (blue), $T_{31}$ (green) and $T_{32}$ (red) as a function of Fermi energy $E_F$ for (a) $t_{12}=-0.25$ and (b) $t_{12}=-1$. For all cases: $M_1=M_2=M_3=4$, $\epsilon_C=0$, $\epsilon_{i}=0$. Threshold energy $E_F^{th1}=-2.62$, $E_F^{th2}=-1.48$, $E_F^{th3}=-0.48$.}\label{fig9}
\end{figure}

Let us now consider the perpendicular junction (Fig.\ref{fig2}c) for which the coupling  region is reduced to a single strip. The transport characteristics are presented in Fig.\ref{fig9}. For the bend configuration and for a weak coupling $t_{12}=-0.25$ to the floating electrode 2 (Fig.\ref{fig9}a) the conductance $\mathcal{G}_{13,2}$ and the voltage $V_2$ are suppressed because of small values of transmissions $T_{31}$ and $T_{21}$ (see Eq.\ref{eq6} and Eq.\ref{eq7} respectively). We can notice that for both these situations, weak and strong coupling, dips in the conductance and transmission curves disappeared (compare plots in Fig.\ref{fig4} and Fig.\ref{fig8}). This happens for the simple reason that the waves propagate straight from the source to the drain electrode. Mode matching problem does not exist in this case and that bend resistance can no longer be observed. Comparing these results with those for the planar configuration one can quantitatively determine bend resistance in transport.

\section{Non-linear response regime} \label{sec:4}

Recently Xu \cite{Xu1} and Csontos and Xu \cite{Xu2} examined theoretically electron Y-branch switch device (YBS) in non-linear response. Using scattering-matrix method, they \linebreak showed that when finite voltages are applied to the left and right branches in the push-pull fashion ($V_l=V/2$ and $V_r=-V/2$), the output voltage $V_c$ in the central electrode is generally negative. They showed also that for a weak nonlinear response YBS exhibits a parabolic behaviour for $V_c$ at low temperatures:
\begin{eqnarray}\label{eq18}
V_c=-\frac{1}{8}\alpha V^{2}+\mathcal{O}(V^{4}),
\end{eqnarray}
where
\begin{eqnarray}\label{eq19}
\alpha=e\frac{\partial T_{cl}(E_{F})/\partial E_F}{T_{cl}(E_F)}
\end{eqnarray}
and $T_{cl}(E)$ is a transmission coefficient between the central and the left electrode. The result was obtained for a symmetric situation $T_{cl}=T_{cr}$. Generally, the coefficient $\alpha>0$, because $T_{cl}$ increases with $E_F$. The formula holds also for higher temperatures that is why one has to include blurring of the step of the Fermi function. The coefficient $\alpha$ is given by an average value of the transmission and its derivative.

\begin{figure}[ht]
\includegraphics[width=0.5\textwidth]{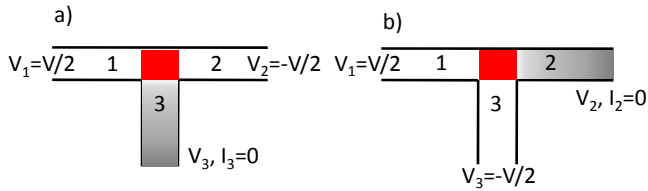}
\caption{(Color online) Planar device configurations for non-linear regime. a) Straight configuration. First and the second electrode is biased in push-pull configuration and the 3rd one is floating one (detector, shaded area). b) Bend configuration. First and third electrode is source and drain and 2nd one is floating (shaded area). The coupling region is marked as the red square in the centre of the device.}\label{fig10}
\end{figure}

In our paper we would like to present similar effects for the T-shape ballistic junction for the straight and bend configuration (see Fig.\ref{fig10}). The source and the drain electrodes are biased in push-pull fashion: $V_1=V/2$, $V_2=-V/2$ for the straight (see Fig.\ref{fig10}a) and $V_1=V/2$, $V_3=-V/2$ for the bend configuration (see Fig.\ref{fig10}b), respectively. We expect interesting features for the case with bend resistance, because the derivative $\partial T(E)/\partial E$ can be negative as well as positive in some energy range. The transmission coefficients are calculated according to the procedure described in Sec.II for temperature $T=0$ and assuming that the bias voltage do not shift the bottom of the conduction band.

The chemical potential in the source and drain electrodes is assumed to be $\mu_S=E_F-eV/2$ and $\mu_D=E_F+eV/2$, respectively. We calculate the voltage in the floating electrode from $V_{fl}=(E_F-\mu_{fl})/e$ (for the straight configuration the floating electrode is the third one and for the bend - the second one), where the electrochemical potential in the floating electrode $\mu_{fl}$ can be determined by the following equation:
\begin{eqnarray}\label{eq20}
\frac{2e}{h}\sum_{j=S,D}\int dE\; T_{fl,j}(E_F)[f(E_F-\mu_{fl})-\nonumber\\
-f(E_F-\mu_j)]=0,
\end{eqnarray}
where $f$ is the Fermi distribution function. The right-hand side of the equation is equal to zero because no net current flows through the detector electrode.

\subsection{Straight configuration}

\begin{figure}[ht]
\includegraphics[width=0.5\textwidth]{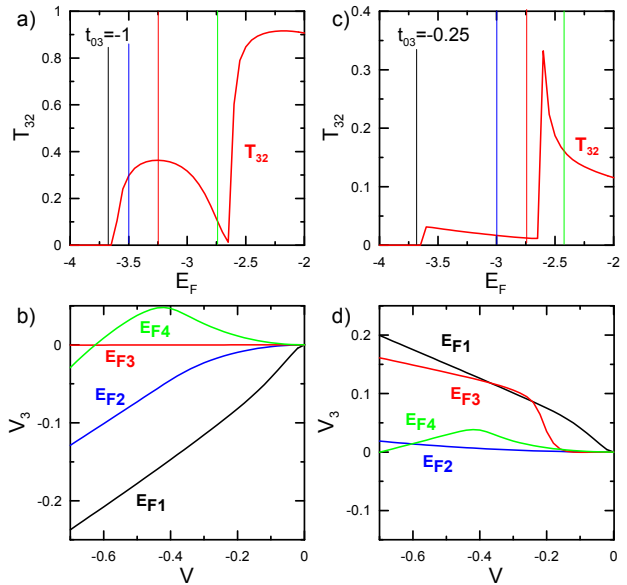}
\caption{(Color online) Results for the straight configuration (with the 3rd electrode left floating): transmission $T_{31}=T_{32}$ (upper panels: a and c) as a function of the Fermi energy $E_F$ and the voltage $V_3$ (lower panels: b and d) in the floating electrode as a function of applied bias voltage $V$. Thin lines in upper panels indicate value of $E_F$ at which the potential $V_3$ was calculated (black - $E_{F1}=-3.7$, blue - $E_{F2}=-3.5$, red - $E_{F3}=-3.25$ and green - $E_{F4}=-2.6$ for the left column and black - $E_{F1}=-3.7$, blue - $E_{F2}=-3$, red - $E_{F3}=-2.75$ and green - $E_{F4}=-2.4$ for the right column). Left side panels (a, b) present the situation for strong coupling to the floating electrode $t_{C3}=-1$), whereas right side panels (c, d) are for weak coupling ($t_{C3}=-0.25$), respectively. For all plots $t_{C1}=t_{C2}=-1$, $\epsilon_{i}=0$ (for $i=1,2,3$), $\epsilon_C=0$ and $M_1=M_2=M_3=4$.}\label{fig11}
\end{figure}

In this section we consider the straight configuration in the device (see Fig.\ref{fig10}a). For the strong coupling of the detector electrode ($t_{C3}=-1$) the transmission coefficient $T_{32}$ is plotted as a function of $E_F$ in Fig.\ref{fig11}a. We have selected four values of the Fermi
energy: $E_{F1}$, $E_{F2}$, $E_{F3}$ and $E_{F4}$ (marked by thin lines in Fig.\ref{fig11}a) for which the derivative $\partial T/\partial E$ have different signs and for which the potential $V_3$ in the floating electrode is determined. The calculated voltage in the floating electrode for the specific energies are plotted below (see Fig.\ref{fig11}b). We show only the negative values of $V$ because the system is symmetrical and yields the same results for the positive values of $V$.

The first value $E_{F1}=-3.7$ lies below the conduction band edge. In this case the  current is exponentially small (due to the thermal excitation of electrons). The voltage curve $V_3$ shows a large slope, which is related to the activation energy. The voltage curve $V_3$ becomes smaller for higher voltages $V$ because in current participate electrons from the conduction band. The second value $E_{F2}=-3.5$ lies in the band close to the pinch-off point. Since the slope of $T_{32}$ is positive the plot of $V_3$ is parabolic (at small $V$) and always negative, in agreement with Xu's suggestions \cite{Xu1,Xu2}. Because $E_{F3}=-3.25$ corresponds to the middle of the conductance step at which $T_{32}$ reaches a maximum (see Fig.\ref{fig11}a), the plot of $V_3$ is zero in a wide range of $V$. The value $E_{F4}=-2.6$ lies in the bend resistance region, where the slope of $T_{32}$ is negative. That is why, the plot of $V_3$ is positive for small $V$. For $V>0.4$ the voltage window is wide enough and the current $I_3$ comprises also contributions from the second conduction channel, where the transmission has a positive slope. Therefore, the $V_3$ curve becomes decreasing for higher bias voltage.

Results for the weak coupling to the detector electrode (for $t_{C3}=-0.25$) are presented in the right column in Fig.\ref{fig11}. The plots for $V_3$ are now different than for the case of strong coupling (compare Fig.\ref{fig11}d and b). For the first value $E_{F1}$ the current is exponentially small. In this case $V_3$ is positive (see Fig.\ref{fig11}d) due to negative slope of the transmission curves $T_{32}=T_{31}$. For $E_{F2}=-3$ the transmission coefficient $T_{32}$ is monotonically decreasing function of $E_F$ and the voltage $V_3$ slowly increases with the bias $V$. The third value $E_{F3}-2.75$ lies below the second transmission step. The potential $V_3$ is negative for small bias $V$ and it becomes large positive when the voltage window is large enough to reach the second transmission step (at $V>0.2$). The large increase of $V_3$ is due to the Wigner threshold effect (see the large peak in $T_{32}$ in Fig.\ref{fig11}c). $E_{F4}=-2.4$ lies above new transmission step where the slope of the transmission curve is negative, thus the voltage $V_3$ in the floating electrode is positive.

As we showed previously the voltage in the detector electrode can be both positive  and negative due to changes of $\partial T(E)/\partial E$. Changing the energy $E_F$ in the whole device can cause oscillations of the voltage in the detector electrode due to interference effects that appear in the T-shaped ballistic junctions. The shape and the amplitude of the oscillations depend on the transmission coefficients.

\begin{figure}[ht]
\includegraphics[width=0.5\textwidth]{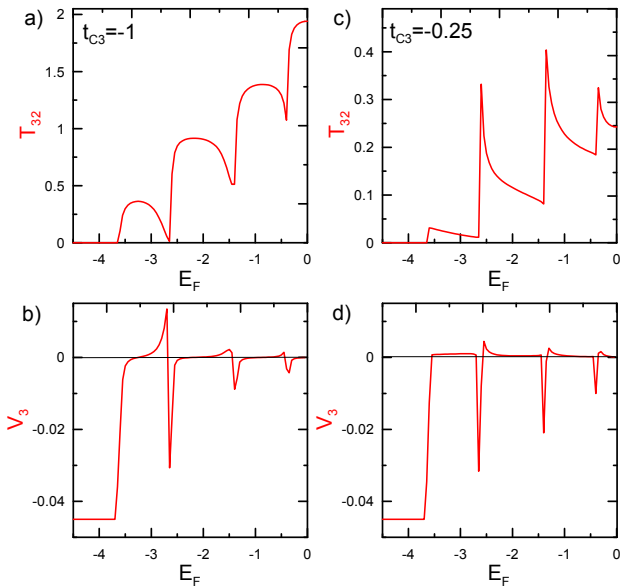}
\caption{(Color online) Results for the straight configuration. Transmission coefficients  $T_{32}=T_{31}$ (upper panels) and voltage $V_3$ in the floating electrode (lower panels) as a function of Fermi energy $E_F$ for: a), b) strong coupling of the floating electrode ($t_{C3}=-1$) and c), d) for weak coupling of the floating electrode ($t_{C3}=-0.25$). For all cases $\epsilon_{i}=0$ (for $i=1,2,3$), $\epsilon_C=0$, $t_{C1}=t_{C2}=-1$, $M_1=M_3=4$, $V=-0.1$.}\label{fig12}
\end{figure}

In Fig.\ref{fig12} the voltage $V_3$ in the floating electrode is calculated as a function of $E_F$ for two cases: strong coupling (see Fig.\ref{fig12}a, b) and weak coupling of the floating electrode (see Fig.\ref{fig12}c, d). The transmission coefficients are presented in the upper panels. For $t_{C3}=-1$ (see Fig.\ref{fig12}b) we observe the appearance of the sharp peaks in $V_3(E_F)$ characteristics due to bend resistance. Large dips appear whenever new transmission channel opens because $\partial T_{32}/\partial E$ is negative. For weak coupling one can notice that the voltage $V_3$ is small and positive, but on the other hand we observe large negative dips caused by the Wigner cusps in $T_{32}$.

\subsection{Bend configuration}

\begin{figure}[ht]
\includegraphics[width=0.5\textwidth]{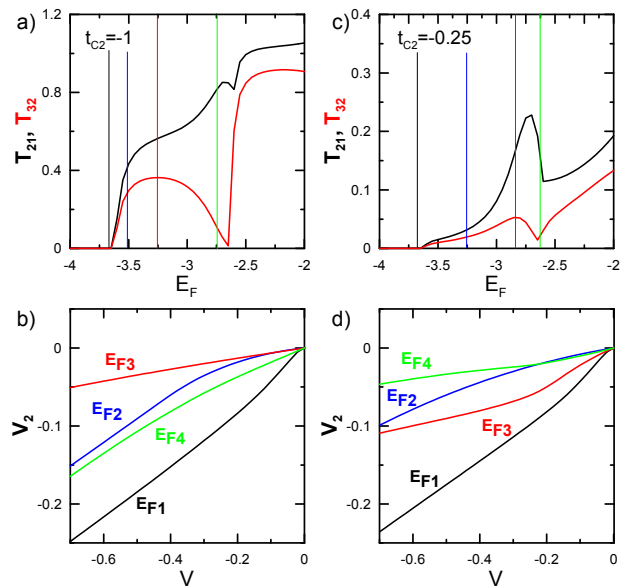}
\caption{(Color online) Results for the bend configuration (with the 2nd electrode left floating): transmission $T_{32}$ - red curve, $T_{21}$ - black curve (upper panels: a and c) as a function of the Fermi energy $E_F$ and the voltage $V_2$ (lower panels: b and d) in the floating electrode as a
function of applied bias voltage $V$. Thin lines in upper panels indicate value of $E_F$ at which the potential $V_2$ was calculated (black - $E_{F1}=-3.7$, blue - $E_{F2}=-3.55$, red - $E_{F3}=-3.25$ and green - $E_{F4}=-2.75$ for the left column and black - $E_{F1}=-3.7$, blue - $E_{F2}=-3.25$, red - $E_{F3}=-2.8$ and green - $E_{F4}=-2.6$ for the right column). Left side panels (a, b) present the situation for strong coupling to the floating electrode $t_{C2}=-1$), right side panels (c, d) are for weak coupling ($t_{C2}=-0.25$), respectively. For all plots $t_{C1}=t_{C3}=-1$, $\epsilon_{i}=0$ (for $i=1,2,3$), $\epsilon_{C}=0$ and $M_1=M_2=M_3=4$.}\label{fig13}
\end{figure}

Fig.\ref{fig13} presents the results for the bend configuration \linebreak (Fig.\ref{fig10}b). Description used in the previous configuration can be applied to this one, even when the transmission coefficients $T_{21}$ and $T_{32}$ have different energy dependencies. For this configuration we see that $V_2$ in the floating electrode is always negative both for strong (see Fig.\ref{fig13}b) and weak (see Fig.\ref{fig13}d) coupling of the floating electrode for whole range of Fermi energies (for regions with bend resistance as well as the Wigner threshold effect). This behaviour originates from competition between currents in the direct and side channels. When the transmission $T_{32}$ to the side channel shows the bend resistance (in the vicinity of $E_F=-2.75$), the transmission $T_{21}$ through the direct channel increases significantly because of flux conservation in the system and filtering properties of the side channel.

\begin{figure}[ht]
\includegraphics[width=0.5\textwidth]{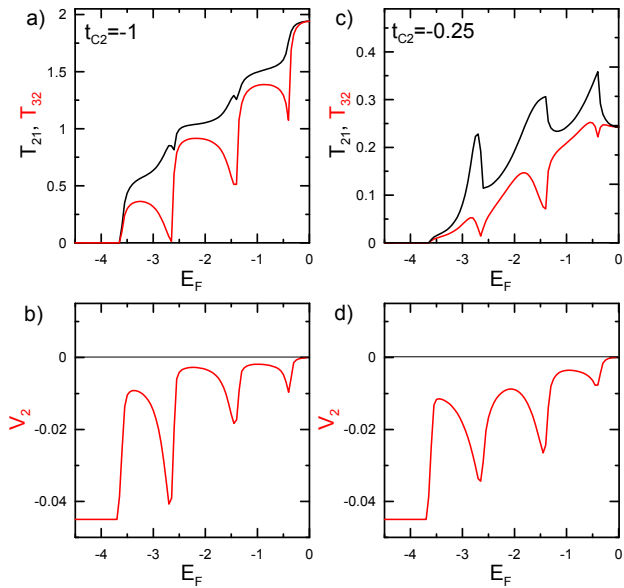}
\caption{(Color online) Results for the bend configuration. Transmission coefficients
$T_{32}$ - red curve, $T_{21}$ - black curve (upper panels) and voltage $V_2$ in the floating
electrode (lower panels) as a function of Fermi energy $E_F$ for: a), b)
strong coupling of the floating electrode ($t_{C2}=-1$), c), d) for weak coupling
of the floating electrode ($t_{C2}=-0.25$). For all cases $\epsilon_{i}=0$
(for $i=1,2,3$), $\epsilon_C=0$, $t_{C1}=t_{C3}=-1$, $M_1=M_3=4$, $V=-0.1$.}\label{fig14}
\end{figure}

Fig.\ref{fig14} presents the voltage $V_2$ in the floating electrode for the whole range of Fermi energy. For both situations: the strong (see Fig.\ref{fig14}a, b) and for the weak coupling to the floating electrode (see Fig.\ref{fig14}c, d) the voltage $V_2$ remains negative. The dips appear when the transmission $T_{21}$ have positive slope even for the bend resistance region where the $\partial T_{32}/\partial E$ yields positive values.

\section{Summary} \label{sec:5}

In this paper we have presented transport studies for the T-shaped three-terminal ballistic junction in the linear and non-linear regime. In the model we omitted electron-electron interactions, which seem to play a minor role for the small voltage regime \cite{Butt2}. Using the tight-binding approach, the Green's function method and the multi-probe Landauer-B\"{u}ttiker formalism, we have performed numerical calculations of transport characteristics for different model parameters. The results have been analyzed for the straight and the bend configuration of the applied bias voltages.

In the linear regime we have analyzed three interference effects: the Wigner threshold  effect, filtering of the electrons and the bend resistance. For the straight configuration the conductance characteristics present the different shapes of the Wigner singularity depending on the coupling between the leads and the coupling region. For equal couplings we obtain the saddle-like singularity in the conductance. For strong couplings to the floating electrode and weak to the drain electrode - the downwarded cusp has been seen, whereas the upwarded cusp has appeared for small couplings. It was shown that the floating electrode filters electron waves in the main channel. The filtering process depends on the length of the transverse-directed wave vector. In the bend configuration we have studied the mod matching and its influence on the conductance in the direct channel and the voltage in the floating electrode. Along with increasing the transverse wave vector length we observed wide dips in the conductance characteristics, which resulted by bend resistance. As a result we observed peaks in the voltage characteristics. The back-action was observed in experiments \cite{Ram} as a distinct change of the voltage in the floating electrode. A recent experiment performed by Wrobel {\it et al.} \cite{Wrobel} revealed periodic oscillations of the voltage in the floating electrode, which was apparently back-action connected with filtering and bend resistance.

For the non-linear regime we have analyzed the influence of interference effects in the voltage in the floating electrode. According to Xu \cite{Xu3,Xu1,Xu2} the voltage of the floating electrode is in the most cases negative and exhibits a parabolic behaviour. We showed that the back-action voltage can be positive in the regions where the bend resistance and threshold effects were relevant. This effect was observed for the straight configuration only. Moreover, for the fixed source-drain voltage we showed that the Wigner threshold effect and the bend resistance can be observed in the voltage characteristics due to back actions in the floating electrode. For the bend configuration we observed voltage oscillations on the floating electrode which decreased with decreasing coupling between floating electrode and central region of the device.

\section{Acknowledgments} \label{sec:5}

This work was supported in part by the research project of Ministry of Science and Higher Education (Poland) No. N202/229437 and No. N202/103936 as well as by the EU project "Marie Curie ITN NanoCTM".


\begin{thebibliography}{99}
\bibitem{Reed} M.A. Reed, W.P. Kirk, {\it Nanostructure Physics and Fabrication}, (Academic, New York, 1989) and C.W.J. Beenakker and H. van
Houten, in {\it Solid state Physics: Advances in Research and Applications}, edited by H. Ehrenreich and D. Turnbull, (Academic, New York, 1991), Vol. 44, pp. 1-228
\bibitem{Fano} U. Fano, Phys. Rev. {\bf124}, 1866 (1961)
\bibitem{Tekman} E. Tekman, P.F. Bagwell, Phys. Rev. B {\bf48}, 2553 (1993)
\bibitem{Stone} J.U. N\"{o}ckel, A.D. Stone, Phys. Rev. B {\bf50}, 17415 (1994).
\bibitem{Kobayashi} K. Kobayashi, H. Aikawa, A. Sano, S. Katsumoto, Y. Iye, Phys. Rev. B {\bf70}, 035319 (2004)
\bibitem{Wu} J.C. Wu, M.N. Wybourne, W. Yindeepol, A. Weisshaar, S.M. Goodnick, Appl. Phys. Lett. {\bf59}, 102 (1991)
\bibitem{Weisshaar} A. Weisshaar, J. Lary, S.M. Goodnick, V.K. Tripathi, J. Appl. Phys. {\bf70}, 335 (1991)
\bibitem{Laux} S.E. Laux, A. Kumar, M.V. Fischetti, J. Appl. Phys. 95, 5545 (2004)
\bibitem{Baranger} H.U. Baranger, Phys. Rev. B {\bf42}, 11479 (1990).
\bibitem{Schult} R.L. Schult, D.G. Ravenhall, H.W. Wyld, Phys. Rev. B {\bf39}, 5476 (1989); D.G. Ravenhall, H.W. Wyld, R.L. Schult, Phys. Rev. Lett. {\bf62}, 1780 (1989)
\bibitem{Chen} Y.P. Chen, X.H. Yan, Y.E. Xie, Phys. Rev. B {\bf71}, 245335 (2005)
\bibitem{Worschech} L. Worschech, B. Weidner, S. Reitzenstein, A. Forchel, Appl. Phys. Lett. {\bf78}, 3325 (2001)
\bibitem{Wigner} E.P. Wigner, Phys. Rev. {\bf73}, 1002 (1948); see also: R.L. Landau, E.M. Lifshitz, in {\it Quantum Mechanics}, 3rd ed. (Pergamon, Oxford, 1977), Chap. 18; R.G. Newton, in {\it Scattering Theory of Waves and Particles}, (Springer-Verlag, New York, 1982), Chap. 17.2
\bibitem{wignereffect} L.M. Delves, Nucl. Phys. {\bf8}, 358 (1958); H. Hotop, M.W. Ruf, I.I. Fabrikant, Phys. Scr. {\bf110}, 22 (2004); J. Weiner, V.S. Bagnato, S. Zilio, P.S. Julienne, Rev. Mod. Phys. {\bf71}, 1 (1999)
\bibitem{Bulka} B.R. Bu{\l}ka, A. Tagliacozzo, Phys. Rev. B {\bf79}, 075436 (2009)
\bibitem{Ram} A. Ramamoorthy, J.P. Bird, J.L. Reno, J. Phys.: Condens. Matter {\bf19}, 276205 (2007)
\bibitem{Shorubalko} I. Shorubalko, H.Q. Xu, I. Maximov, D. Nilsson, P. Omling, L. Samuelson, W. Seifert, IEEE Electron Device Lett. {\bf23}, 377 (2002)
\bibitem{Xu2002} H.Q. Xu, Appl. Phys. Lett. {\bf80}, 853 (2002)
\bibitem{Papadopoulos} C. Papadopoulos, A. Rakitin, J. Li, A.S. Vedeneev, J.M. Xu, Phys. Rev. Lett. {\bf85}, 3476 (2000)
\bibitem{Bandaru} P.R. Bandaru, C. Daraio, S. Jin, A.M. Rao, Nat. Mater. {\bf4}, 663 (2005)
\bibitem{Xu3} H. Xu, Nat. Mater. {\bf 4}, 649 (2005)
\bibitem{Brat} P. Samuelson, A. Brataas, Phys. Rev. B {\bf81}, 184422 (2010)
\bibitem{Xu1} H.Q. Xu, Appl. Phys. Lett. {\bf78}, 2064 (2001)
\bibitem{Scho} I. Shorubalko, H.Q. Xu, I. Maximov, P. Omling, W. Seifert, Appl. Phys.  Lett. {\bf79}, 1384 (2001)
\bibitem{Xu2} D. Csontos and H.Q. Xu, Phys. Rev. B {\bf67}, 235322 (2003)
\bibitem{Butt1} M. B\"{u}ttiker, Phys. Rev. Lett. {\bf57}, 1761 (1986)
\bibitem{Fisher} D.S. Fisher and P.A. Lee, Phys. Rev. B {\bf23}, 6851 (1981)
\bibitem{Ando} T. Ando, Phys. Rev. B {\bf44}, 8017 (1991)
\bibitem{Knop} M. Knop, M. Richter, R. Ma{\ss}mann, U. Wieser, U. Kunze, D. Reuter, C. Riedesel, A.D. Wieck, Semicond. Sci. Technol. {\bf20}, 814 (2005)
\bibitem{Wees} B.J. van Wees, Phys. Rev. B {\bf43}, 12431 (1991)
\bibitem{Butt2} For bias voltages comparable to the Fermi energy electron interactions can be relevant and lead to opening of a gap in the transport     window - see: A.N. Jordan and M. B\"{u}ttiker, Phys. Rev. B {\bf77}, 075334 (2008)
\bibitem{Wrobel} J. Wr\'{o}bel, P. Zagrajek, M. Czapkiewicz, M. Bek, D. Sztenkiel, K. Fronc, R. Hey, K.H. Ploog, B.R. Bu{\l}ka, Phys. Rev. B {\bf81},
    233306 (2010)
\end{thebibliography}
\end{document}